\documentclass[preprint2]{aastex}
\usepackage{amsmath}
\usepackage{multirow}
\usepackage[T1]{fontenc}
\pdfpageattr{/Group <</S /Transparency /I true /CS /DeviceRGB>>}
\graphicspath{{figures/}}
\defcitealias{Boue_Laskar_Icarus_2006}{BL06}
\defcitealias{Boue_Laskar_Icarus_2009}{BL09}
\defcitealias{Kaib_etal_ApJ_2011}{K11}
\renewcommand{\vec}[1]{{\boldsymbol #1}}
\newcommand{\lap}[2]{{\rm b}_{#1}^{(#2)}}

\def\abs#1{\left\vert#1\right\vert}
\def\norm#1{\left\Vert#1\right\Vert}

\newcommand{\grad}[1]{\boldsymbol{\nabla}_{\!#1}}
\newcommand{\moy}[2]{\left\langle{#2}\right\rangle_{#1}}
\def\crm{\cr\noalign{\medskip}}

\def\m@th{\mathsurround=0pt}
\def\EQM#1{\vcenter{\normalbaselines\m@th
    \ialign{${\displaystyle ##}$\hfil&&\ ${\displaystyle ##}$\hfil\crcr
    \mathstrut\crcr\noalign{\kern-\baselineskip}
    \noalign{\smallskip}
    #1\crcr\mathstrut\crcr\noalign{\kern-\baselineskip}}}}

\newcommand{\be}{\begin{equation}}
\newcommand{\ee}{\end{equation}}

\def\Dron#1#2{\frac{\partial#1}{\partial#2}}

\newcommand{\bpm}{\begin{pmatrix}}
\newcommand{\epm}{\end{pmatrix}}
\newcommand{\figLeVerrier}{
\begin{figure}[t]
\plotone{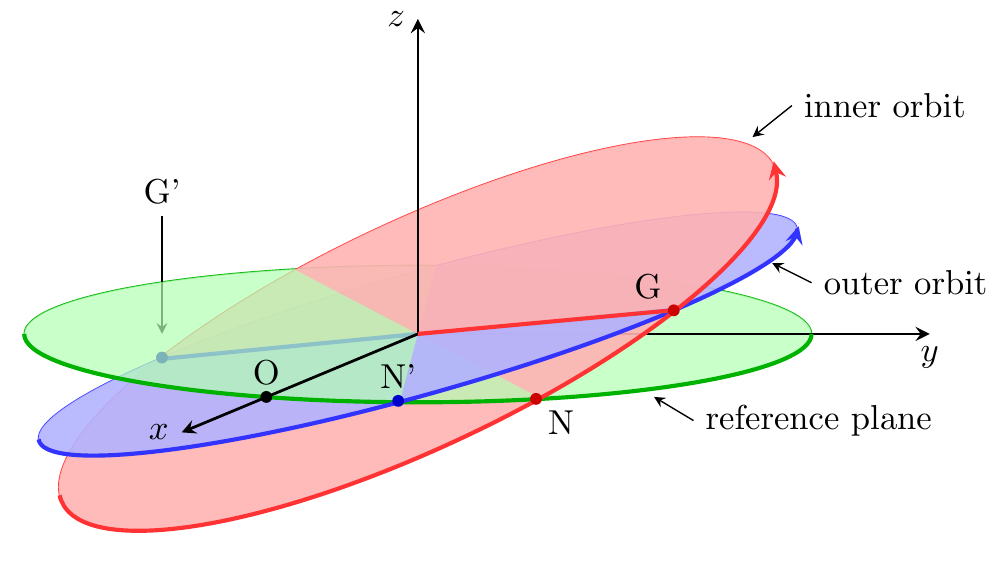}
\caption{\label{fig.LeVerrier} Orbit orientation. ON'N defines the
reference plane. O is the origin of longitudes, N and N' are the
ascending nodes of the orbits of $m$ and $m'$ relative to the reference
plane, respectively. G and G' are the ascending nodes of the 
orbit $m$ and $m'$ relative to the orbit $m'$ and $m$,
respectively.}
\end{figure}
}
\newcommand{\figCompa}{
\begin{figure}[t]
\plotone{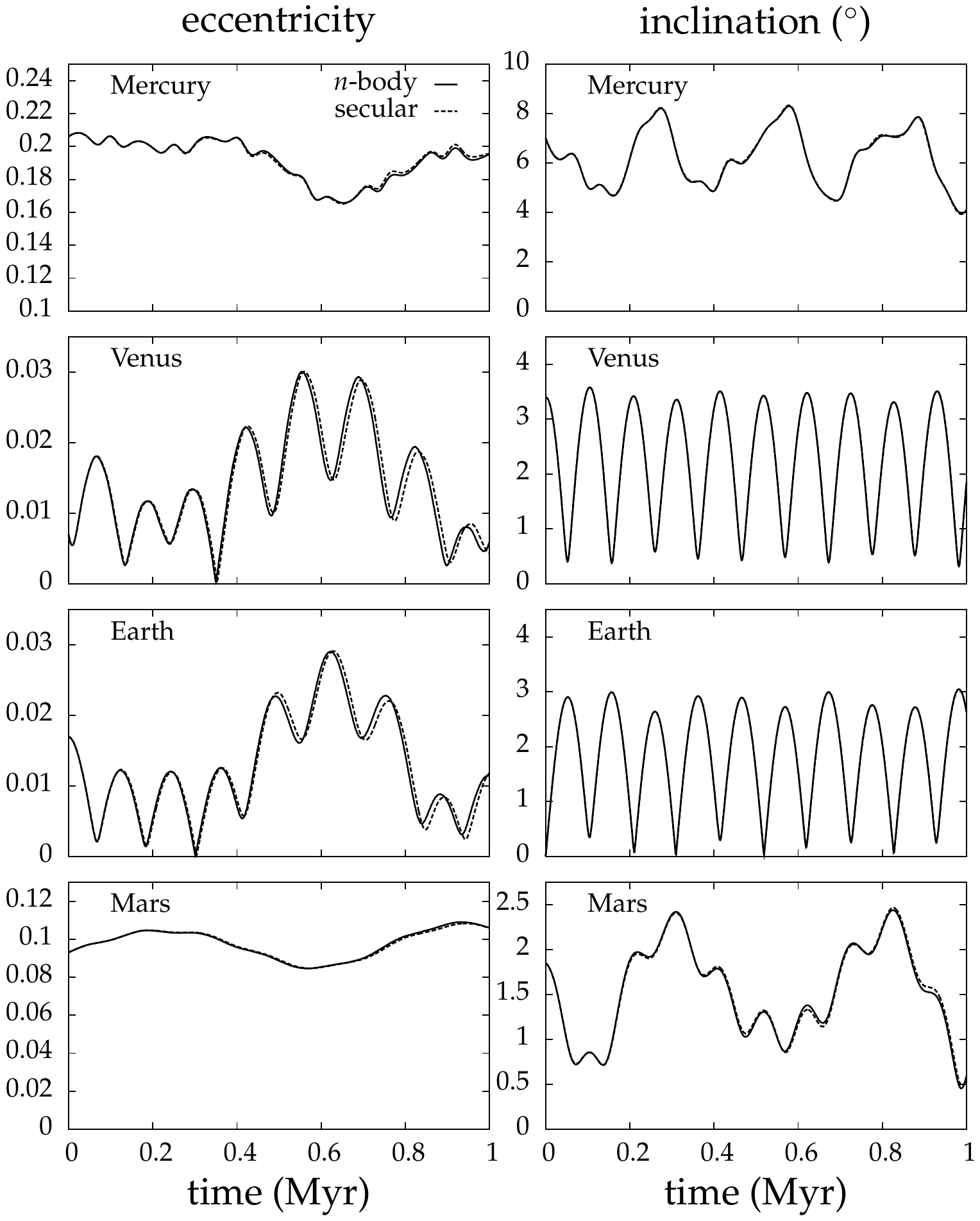}
\caption{\label{fig.comparison1} Comparison between an $n$-body
integration and a secular integration on a system composed of the four
inner planets of our solar system followed over one million years.}
\end{figure}
}
\newcommand{\figCompb}{
\begin{figure}[t]
\plotone{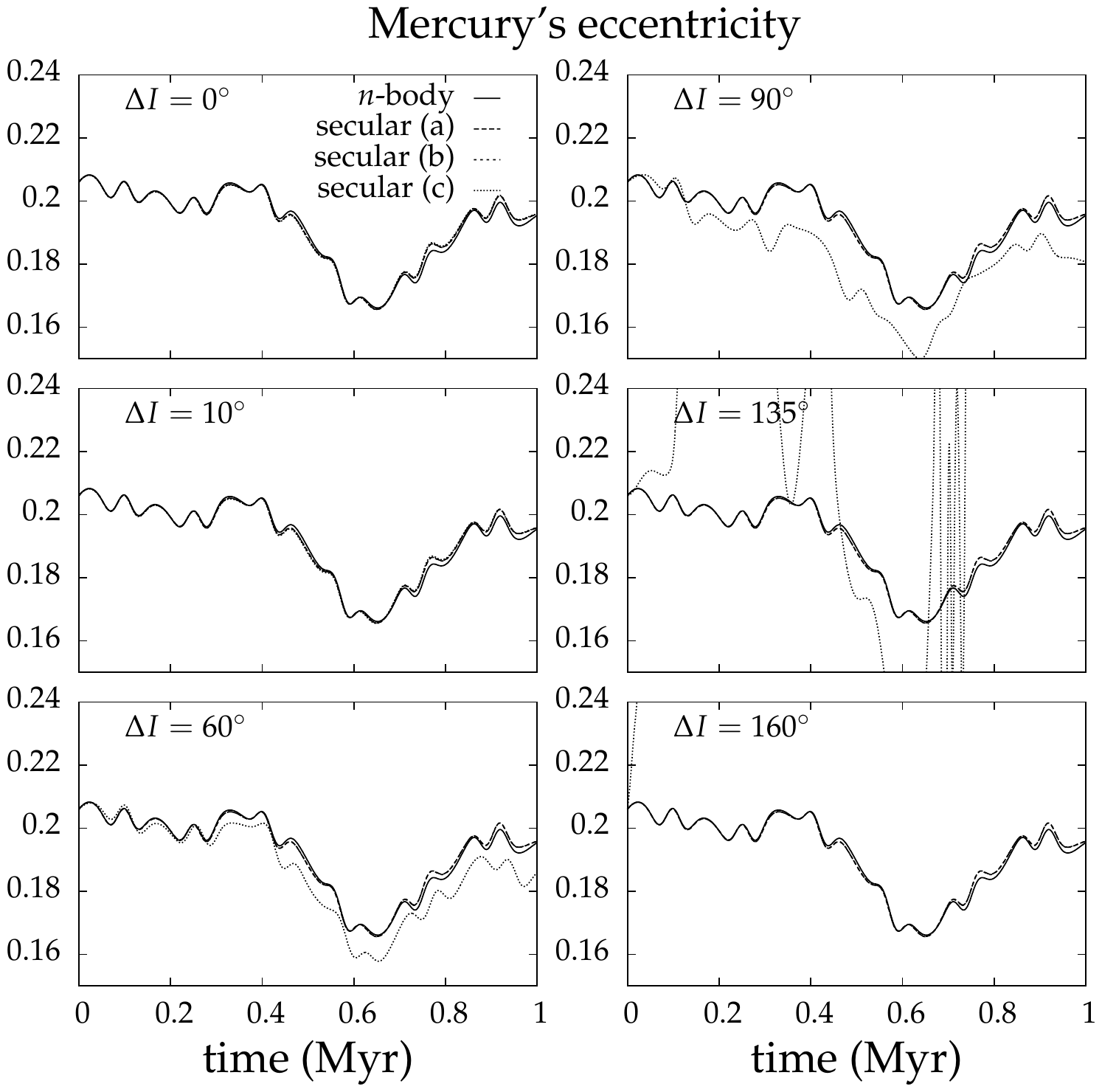}
\caption{\label{fig.comparison2} Effect of a rotation by $\Delta I$ of
the whole system on Mercury's eccentricity simulated with
different secular models. The latter are: (a) expansion in
eccentricity and mutual inclination in Milankovitch's variables, (b)
expansion in eccentricity and mutual inclination using Le~Verrier's
equations, and (c) expansion in eccentricity and {\em absolute}
inclination \citep{Laskar_Robutel_CeMDA_1995}. Models (a) and (b)
overlap and cannot be distinguished from one another.}
\end{figure}
}
\newcommand{\figCompbb}{
\begin{figure}[t]
\plotone{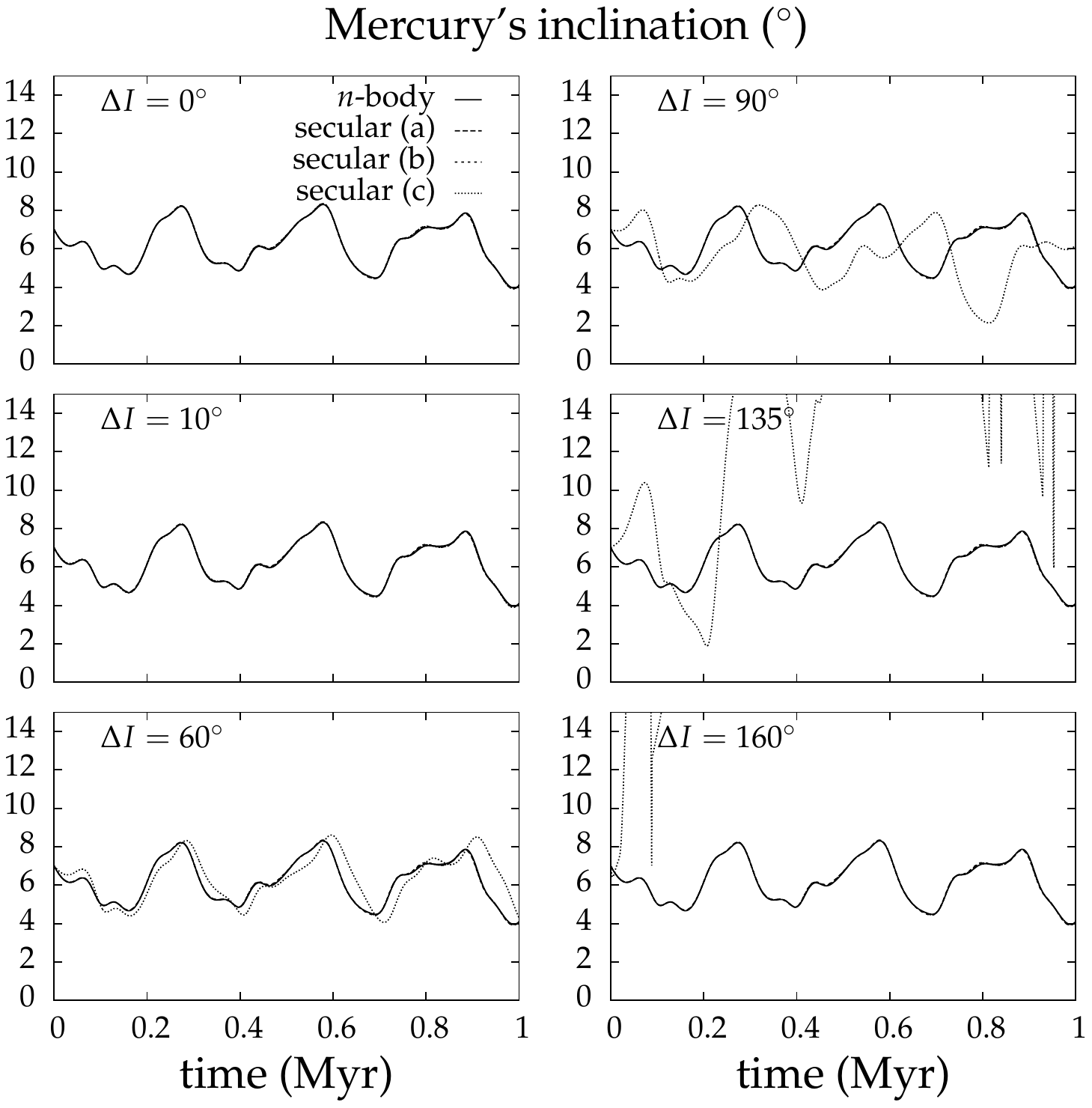}
\caption{\label{fig.comparison2b} Same as Fig.~\ref{fig.comparison2} but
with Mercury's inclination. In all panels, inclinations are computed
with respect to the same reference plane after performing a rotation of
$-\Delta I$ on the output of each simulation. Models (a) and (b) are
overlapping and cannot be distinguished from one another.}
\end{figure}
}
\newcommand{\figCompc}{
\begin{figure}[t]
\plotone{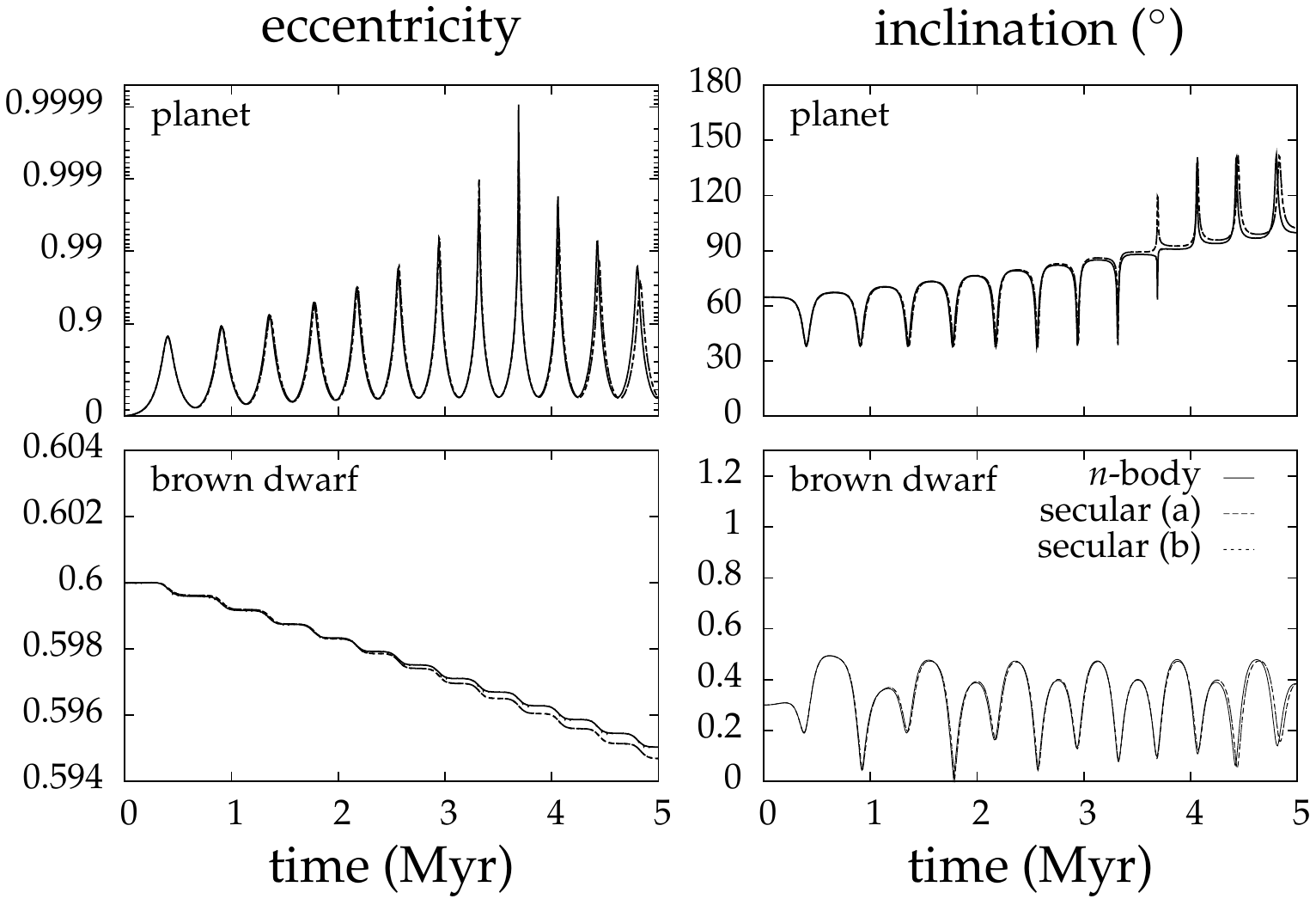}
\caption{\label{fig.comparison3} Comparison between an $n$-body
simulation and secular integrations of a hierarchical three-body system.
The parameters are taken from \citep[fig.~1]{Naoz_etal_nature_2011}.
The system is composed of a planet with mass $m_1=1 M_J$, at $a_1 = 6$ au with
an initial eccentricity of $e_1=0.001$, and a brown dwarf with mass $m_2
= 40 M_J$, at $a_2 = 100$ au and an initial eccentricity $e_2=0.6$. The
initial mutual inclination is $J=65^\circ$. The secular
models are: (a) expansion in eccentricity and mutual inclination in
Milankovitch's variables and (b) expansion in eccentricity and mutual
inclination using Le~Verrier's equations. Note that the eccentricity of
the planet (upper left panel) is plotted in a $\tanh^{-1}$ scale.}
\end{figure}
}

\newcommand{\Tabci}{
\begin{table}
\begin{center}
\caption{\label{tab.ci}Coefficients of the secular expansion of the
perturbing function in eccentricity and mutual inclination.}
\begin{tabular}{l} \hline
$\displaystyle
\EQM{
c_1 &= \frac{1}{2}\lap{1/2}{0} \crm
c_2 &= \frac{1}{4}\alpha\lap{3/2}{1} \crm 
c_3 &= \frac{3}{4}\alpha\lap{3/2}{0} - \frac{1}{2}(1+\alpha^2)\lap{3/2}{1} \crm
c_4 &= \frac{9}{16}\alpha^2\lap{5/2}{0} \crm
c_5 &= \frac{9}{32}\alpha\lap{5/2}{1} \crm
c_6 &= \frac{45}{32}\alpha^2\lap{5/2}{0} 
     - \frac{9}{16}\alpha(1+\alpha^2)\lap{5/2}{1} \crm
c_7 &= \frac{21}{32}\alpha^2\lap{5/2}{0}
     - \frac{3}{16}\alpha(1+\alpha^2)\lap{5/2}{1} \crm
c_8 &= \frac{15}{32}\alpha^2\lap{5/2}{0}
     - \frac{3}{16}\alpha(3+\alpha^2)\lap{5/2}{1} \crm
c_9 &= \frac{15}{32}\alpha^2\lap{5/2}{0}
     - \frac{3}{16}\alpha(1+3\alpha^2)\lap{5/2}{1} \crm
c_{10}&=-\frac{15}{16}\alpha\lap{5/2}{0}
       + \frac{3}{8}\lap{5/2}{1} \crm
c_{11}&=-\frac{15}{16}\alpha^2\lap{5/2}{0}
       + \frac{3}{8}\alpha^3\lap{5/2}{1} \crm
c_{12}&= \frac{15}{8}\alpha(1+\alpha^2)\lap{5/2}{0}
       - \frac{3}{16}(4+9\alpha^2+4\alpha^4)\lap{5/2}{1} \crm
}
$ \\ \hline
\end{tabular}
\end{center}
\end{table}
}
\newcommand{\Tabnotation}{
\begin{table*}
\begin{center}
\caption{\label{tab.notation}Notation.}
\begin{tabular}{clll} \\\hline
& variable & Ref. & description \\ \hline \hline
\multirow{28}{*}{\rotatebox{90}{planetary orbital and physical elements}}
& $\vec r$, $\vec r'$ & & astrocentric position \\
& $\tilde{\vec r}$, $\tilde{\vec r'}$ & & barycentric velocity \\
& $a$, $a'$ & & semimajor axis \\
& $\alpha$ & & semimajor axis ratio $a/a'$ \\
& $\lambda$, $\lambda'$ & & mean longitude \\
& $e$, $e'$ & & eccentricity \\
& $\omega$, $\omega'$ & & argument of pericenter \\
& $\varpi$, $\varpi'$ & & longitude of pericenter \\
& $I$, $I'$ & & absolute inclination \\
& $J$ & & mutual inclination between two planets \\
& $\rho$ & & $\sin(J/2)$ \\
& $\sigma$ & & $\cos(J/2)$ \\
& $\Omega$, $\Omega'$ & & longitude of the ascending node $\rm ON$ \\
& $\tau$ & & $\rm ON + NG$ \\
& $\rm O$ & Fig.~\ref{fig.LeVerrier} & origin of longitude \\
& $\rm N$, $\rm N'$ & Fig.~\ref{fig.LeVerrier} & ascending node relative to the reference plane \\
& $\rm G$, $\rm G'$ & Fig.~\ref{fig.LeVerrier} & ascending node relative to another orbit plane \\
& $\Delta$ & & mutual distance $\norm{\vec r-\vec r'}$ \\
& $m$, $m'$ & & planet mass \\
& $\beta$, $\beta'$ & & reduced mass \\
& $\mu$, $\mu'$ & & ${\cal G}(m_0+m)$  \\
& $\Lambda$, $\Lambda'$ & & $\beta\sqrt{\mu a}$ \\
& $\vec e$, $\vec e'$ & & eccentricity vector \\
& $\vec j$, $\vec j'$ & & dimensionless angular momentum $\sqrt{1-e^2}\vec w$ \\
& $\vec w$, $\vec w'$ & & unit vector along the orbital angular momentum \\
& $\vec \xi$, $\vec \xi'$  & Eq.~(\ref{eq.varSouriau}) & Souriau variable $\vec j + \vec e$ \\
& $\vec \eta$, $\vec \eta'$ & Eq.~(\ref{eq.varSouriau}) & Souriau variable $\vec j - \vec e$ \\
& $T_j$, $V_j$, $W_j$ & Eq.~(\ref{eq.Abdullah}) & Abdullah variables quadratic in inclination and eccentricity \\
\hline
\multirow{8}{*}{\rotatebox{90}{stellar parameters}}
& $m_0$ & & star mass \\
& $R_0$ & & star radius \\ 
& $C$ & & moment of inertia along the short axis \\ 
& $k_2$ & & second fluid Love number \\
& $J_2$ & Eq.~(\ref{eq.Jom}) & quadrupole gravitational harmonic \\
& $\omega_0$ & & rotation vector \\
& $\vec s$ & & spin axis \\
& $\vec L$ & & angular momentum $C\omega_0\vec s$ \\
\hline
\multirow{9}{*}{\rotatebox{90}{Hamiltonians}}
& $H_{\rm close}$ & Eq.~(\ref{eq.Hclose}) & Hamiltonian of packed system \\
& $\bar H_{\rm close}$ & Eq.~(\ref{eq.HamPP}) & secular Hamiltonian of packed system \\
& $\bar H_{\rm hierar}$ & Eq.~(\ref{eq.HamPC}) & secular Hamiltonian of hierarchical system \\
& $\bar H_{\rm spin}$ & Eq.~(\ref{eq.HamSP}) & secular Hamiltonian of spin-orbit interaction \\
& $\bar H_{\rm relat}$ & Eq.~(\ref{eq.HamRG}) & secular Hamiltonian of general relativity \\
& $\lap{s}{k}(\alpha)$ & Eq.~(\ref{eq.Laplace}) & Laplace coefficient \\
& $f_j$ & Eq.~(\ref{eq.LeVerriera}) & coefficients of Le Verrier's expansion of the perturbing function \\ 
& $g_j$ & Eq.~(\ref{eq.g_j}) & coefficients of the perturbing function in Abdullah variables \\
& $c_j$ & Tab.~\ref{tab.ci} & coefficients of the perturbing function in Milankovitch variables \\
\hline
& $\cal G$ & & gravitational constant \\
& $c$ & & speed of light \\
\hline
\end{tabular}
\end{center}
\end{table*}
}
\shorttitle{Perturbed compact planetary systems}
\shortauthors{Bou\'e G. \& Fabrycky D.}
\begin{document}
\title{Compact planetary systems perturbed by an inclined companion: \\I.
Vectorial representation of the secular model}
\author{Gwena\"el Bou\'e\altaffilmark{1,2} and Daniel C. Fabrycky\altaffilmark{1}}
\altaffiltext{1}{Department of Astronomy and Astrophysics, University of Chicago,
5640 South Ellis Avenue, Chicago, IL 60637, USA}
\altaffiltext{2}{Astronomie et Syst\`emes Dynamiques, IMCCE-CNRS UMR 8028,
Observatoire de Paris, UPMC, 77 Av. Denfert-Rochereau, 75014 Paris, France.}
\email{boue@uchicago.edu}
\begin{abstract}
 The non-resonant secular dynamics of compact planetary systems are modeled by a perturbing function which is usually expanded in eccentricity and {\em absolute} inclination with respect to the invariant plane. Here, the expressions are given in a vectorial form which naturally leads to an expansion in eccentricity and {\em mutual} inclination. The two approaches are equivalent in most cases, but the vectorial one is specially designed for those where a quasi-coplanar system tilts as a whole by a large amount. 
Moreover, the vectorial expressions of the Hamiltonian and of the equations of motion are slightly simpler than those given in terms of the usual elliptical elements. 
We also provide the secular perturbing function in vectorial form expanded in semimajor axis ratio
allowing for arbitrary eccentricities and inclinations.
The interaction between the equatorial bulge of a central star and its planets is also provided, as is the relativistic periapse precession of any planet induced by the central star.  We illustrate the use of this representation for following the secular oscillations of the terrestrial planets of the solar system, and for Kozai cycles as may take place in exoplanetary systems. 
\end{abstract}
\section{Introduction}
Observations show nearly half of solar-type stars have a stellar companion \citep{Duquennoy_Mayor_AAP_1991, Raghavan_etal_ApJS_2010} and close-in giant planets are found in such binary systems \citep{Zucker_Mazeh_ApJL_2002, Udry_Santos_ARAA_2007}.  These observations have motivated extensive studies of the dynamics of a single planet evolving in a binary stellar system \citep{Holman_etal_nature_1997, Wu_Murray_ApJ_2003, Fabrycky_Tremaine_ApJ_2007, Wu_etal_ApJ_2007, Correia_etal_CeMDA_2011, Naoz_etal_nature_2011, Lithwick_Naoz_ApJ_2011, Katz_etal_PhRvL_2011, Veras_Tout_MNRAS_2012, Kratter_Perets_ApJ_2012, Naoz_etal_ApJL_2012, Naoz_etal_MNRAS_2013, Naoz_etal_ApJ_2013}. But above all, these systems present interesting dynamics due to the Lidov-Kozai mechanism \citep{Lidov_PSS_1962, Kozai_AJ_1962} which also provides a natural way to form hot Jupiters.  The motion described by Lidov-Kozai dynamics takes place at large inclination and has the structure of a 1:1 resonance between the precession frequency of the longitude of pericenter $\varpi$ and of the longitude of the ascending node $\Omega$ of the planet.  Inside the resonance, the critical angle, equal to the argument of the pericenter $\omega = \varpi -\Omega$, librates around either 90$^\circ$ or 270$^\circ$, and the eccentricity and inclination undergo large amplitude oscillations in antiphase. This behavior combined with tidal dissipation is able to shrink the orbit of a cold Jupiter down to orbital periods of about 3 days. During the evolution, the apsidal precession becomes dominated by general relativity, in which case the system exits the Lidov-Kozai resonance and the orbit evolves through tides as if the companion was nonexistent. The final obliquity of the star with respect to the planet orbit is often quite large \citep{Fabrycky_Tremaine_ApJ_2007, Correia_etal_CeMDA_2011, Naoz_etal_nature_2011, Naoz_etal_ApJL_2012, Li_etal_ApJ_2014, Petrovich_arxiv_2014}. These theoretical predictions are supported by many observations of misaligned systems with a hot Jupiter \citep{Winn_etal_ApJL_2010, Triaud_etal_AA_2010, Triaud_AA_2011, Albrecht_etal_ApJ_2012}.
The dynamical structure associated to the Lidov-Kozai mechanism subsists when the outer stellar companion is replaced by a planet \citep{Terquem_Paploizou_MNRAS_2002, Michtchenko_etal_Icarus_2006, Libert_Henrard_Icarus_2007, Libert_Henrard_CeMDA_2008, Migaszewski_Gozdziewski_MNRAS_2009, Mardling_MNRAS_2010, Naoz_etal_nature_2011, Libert_Delsate_MNRAS_2012}.  The outer planet only needs to be placed on a closer orbit than in the stellar case such that the secular timescale does not exceed the lifetime of the system.  This framework has mainly been used to study the evolution and/or the formation of Jupiters on eccentric orbits.  If both planets form in the same protoplanetary disk, the initial mutual inclination is expected to be small, but it can be generated by high order resonance crossings during planetary migration \citep{Libert_Tsiganis_MNRAS_2009, Libert_Tsiganis_MNRAS_2011}, secular resonance overlaps \citep{Wu_Lithwick_ApJ_2011}, or planet-planet scattering \citep{Nagasawa_Ida_ApJ_2011}. Nevertheless, the latter evolutions are too violent for the Lidov-Kozai mechanism to play a significant role \citep{Beauge_Nesvorny_ApJ_2012}.
In multiplanet systems surrounded by an outer stellar companion, apsidal precession frequencies are dictated by the companion {\em and} by the planet-planet interactions. As a consequence, even at high inclination, if the planet system is sufficiently packed, planet-planet interactions dominate the apsidal motion, the evolution is stabilized with respect to the Lidov-Kozai mechanism, eccentricities remain small, and all planets move in concert \citep{Innanen_etal_AJ_1997, Takeda_etal_ApJ_2008, Saleh_Rasio_ApJ_2009}. These systems are classified as dynamically rigid. Although the Lidov-Kozai evolution is quenched, the planetary mean plane still precesses if it is inclined relative to the orbit of the companion. The long term evolution of such systems can be followed with Laplace-Lagrange second order secular theory. This approach has already been applied in the context of planet formation in a binary system or of satellite formation around Uranus \citep{Batygin_etal_AAP_2011, Batygin_nature_2012, Morbidelli_etal_Icarus_2012, Lai_MNRAS_2014}.  But higher order secular theories that assume absolute inclination remains small for all time can no longer follow the system.
However, the secular dynamics should not simply be followed with Laplace-Lagrange secular model, which predicts quasiperiodic, bounded eccentricities and inclinations. For example, the solar system itself has chaotic secular dynamics \citep{Laskar_nature_1989, Laskar_Icarus_1990}, and only theories accurate to fourth or higher orders in eccentricities and inclinations are able to show the appearance of this chaos \citep{Laskar_these_1984, Lithwick_Wu_ApJ_2011}.
The purpose of this paper is to provide a set of equations describing the secular evolution of non-resonant conservative gravitational systems with a massive central body. The formalism is very general and can be applied to many different types of systems.  In a subsequent paper (Bou\'e and Fabrycky 2014), we use it to study analytically the evolution of the spin-orbit angle in compact planetary systems perturbed by an inclined companion.
Such analyses can a priori be performed numerically using an $n$-body integrator. However, the huge difference between orbital periods and secular timescales makes this approach hardly feasible in a reasonable amount of time. For instance, the 55 Cnc system \citep{Fischer_etal_ApJ_2008}, a compact multiplanet system perturbed by a stellar companion \citep{Mugrauer_etal_AN_2006}, contains five planets whose the innermost has a period of 0.73 days whereas the precession motion of the planetary system is about 70 Myr \citep[][fig.1]{Kaib_etal_ApJ_2011}.  The integration should thus last several hundreds of million years with a time step of a small fraction of a day.
The long term evolution can also be followed with Gauss' method where the equations of motion, averaged over the mean longitudes of all planets, are given by double integrals \citep{Touma_etal_MNRAS_2009}. In practice, the first averaging is analytical while the second has to be computed numerically. This semi-analytic technique is faster than $n$-body codes and it can be applied to a large class of systems, even those with crossing orbits. Indeed, it does not assume any constraints on inclination, eccentricity, nor semimajor axis ratio. 
However, compact systems with large eccentricities or inclinations are likely to be unstable due to resonances overlap. One can thus assume that in most systems which are at least regular over a few secular timescales, each pair of planets is non-exclusively either hierarchical or quasi-coplanar with low eccentricities. These hypotheses are motivated by statistical studies of compact exoplanet systems detected by {\em Kepler} or by radial velocity \citep[e.g.,][]{Tremaine_Dong_AJ_2012, Figueira_etal_AA_2012, Fabrycky_etal_ApJ_2013, Wu_Lithwick_ApJ_2013}. Within this framework, planet-planet interactions are expanded either in semimajor axis ratio, or in mutual inclination and eccentricity. Hence, this technique is more restrictive than Gauss', but both the Hamiltonian and the equations of motion are analytical. This method is thus faster and more appropriate for analytical studies.
 
For this study, the expressions are given in a vectorial form which is independent of any reference frame. This approach, initiated by \citet{Milankovitch_Bull_1939}, has recently been proved very efficient in the study of cometary motion \citep{Breiter_Ratajczak_MNRAS_2005}, of secular spin-orbit evolution and spin-spin interaction \citep{Boue_Laskar_Icarus_2006, Boue_Laskar_Icarus_2009}, of the secular three-body problem \citep{Farago_Laskar_MNRAS_2010, Correia_etal_CeMDA_2011}, and of the secular evolution of satellites \citep{Tremaine_etal_AJ_2009, Tremaine_Yavetz_AJP_2013}. A detailed historical description of the construction of these variables and the associated equations of motion is given in \citet{Rosengren_Scheeres_CeMDA_2014}. The Hamiltonian and the equations of motion, derived in Section~\ref{sec.hamiltonian}, are expressed analytically by means of expansions in eccentricity and mutual inclination on the one hand, and in semimajor axis ratio, on the other. The independence of the vectorial equations from the reference frame is particularly useful for problems where the plane of a planet system tilts by a huge angle with respect to a given reference plane. This specific problem is treated in a subsequent paper (Bou\'e \& Fabrycky, 2014). The model is tested in Section~\ref{sec.numerical} against numerical integrations. The conclusions are given in the last section.
\section{Formalism}
\label{sec.hamiltonian}
\label{sec.notation}
Consider a system composed of an arbitrary number of massive bodies orbiting a central mass $m_0$. For simplicity, the central body is referred to as the central star and the others are called planets. Nevertheless, the formalism is more general and can be applied to other systems. For exemple, it can model stars orbiting a black hole,
or satellite systems.
We utilize Poincar\'e canonical astrocentric variables composed of astrocentric positions and barycentric velocities \citep[e.g.][]{Laskar_Robutel_CeMDA_1995}. Ellipses defined by these variables are not osculating, but for systems with more than three bodies, the formalism is simpler than that involving Jacobi coordinates.  The canonical astrocentric variables (positions and conjugate momenta) of each body are noted $(\vec r, \tilde{\vec r})$, and the orbital elements $(a, \lambda, e, \varpi, I, \Omega)$ represent the semimajor axis, mean longitude, eccentricity, longitude of the pericenter, inclination, and longitude of the ascending node, respectively. All these elements are given with respect to a fixed reference frame. Whenever we consider two planets, quantities associated to the outermost are noted with a prime such as $m'$. We also denote by $J$ the mutual inclination of any pairs of planets. In addition to the previous quantities, we also define $\beta = m_0m/(m_0+m)$, $\mu={\cal G}(m_0+m)$ where ${\cal G}$ is the universal gravitational constant, and $\Lambda=\beta\sqrt{\mu a}$. All quantities are recalled in Tab.~\ref{tab.notation}.
\Tabnotation
\subsection{Close-in interaction}
\label{sec.planetplanet}
This section is devoted to the expansion of the perturbing function in inclination and eccentricity. The typical application of such this approximation is to model the interaction between two planets that are close to each other. In canonical astrocentric variables, the Hamiltonian describing the evolution of a compact pair of planets orbiting a central star
\be
H_{\rm close} = K_{\rm close} + H_{I,\rm close} + H_{D,\rm close}
\label{eq.Hclose}
\ee
is the sum of a Keplerian part
\be
K_{\rm close} = -\frac{\mu\beta}{2a} - \frac{\mu'\beta'}{2a'}\ ,
\ee
an indirect part
\be
H_{I,\rm close} = 
\frac{\tilde{\vec{r}}\cdot \tilde{\vec{r}}'}{m_0}\ ,
\ee
and a direct part
\be
H_{D,\rm close} = - \frac{{\cal G}mm'}{a'}\frac{a'}{\Delta}\ ,
\ee
where $\Delta = \norm{\vec r-\vec r'}$ is the distance between the two planets. Because we only focus on non-resonant secular evolutions, the indirect part $H_{I,\rm close}$ of the Hamiltonian cancels and the Keplerian part $K_{\rm close}$ remains constant. The secular evolution is thus solely controlled by the direct part $H_{D,\rm close}$. 
\subsubsection{Le Verrier's expansion}
The secular component of the expansion of $a'/\Delta$ in eccentricity and {\em absolute} inclination is well known and is expressed analytically in, e.g., \citep{Laskar_Robutel_CeMDA_1995, Ellis_Murray_Icarus_2000}.  Although this approach is very convenient because the resulting expression is a polynomial in the canonical Poincar\'e variables \citep{Laskar_Robutel_CeMDA_1995}, it has not been designed to study quasi-coplanar systems with large absolute inclinations such as compact planetary systems perturbed by a distant and inclined stellar companion.  One needs instead an expansion in {\em mutual} inclination as derived by
\citet{LeVerrier_annales_1855}. The latter is simpler and more compact than expansions in {\em absolute} inclination because it is a special case where one of the absolute inclinations is set to zero.  However, the equations of motion given in {\em mutual} inclination are more cumbersome, especially in systems with more than two planets. In the following, we recall the development of the secular component of $a'/\Delta$ exact up to the fourth order in eccentricity $e, e'$ and mutual inclination $J$.
Then, we recall \citeauthor{LeVerrier_annales_1855}'s equations of
motion. The secular terms of $a'/\Delta$ are
\be
\EQM{
\moy{\lambda,\lambda'}{\frac{a'}{\Delta}} =& &f_1 \crm
&+& f_2\, (e^2+e'^2-4\rho^2) \crm
&+& f_3\, e e' \cos(\omega-\omega') \crm
&+& f_4\, (e^2 e'^2 -4\rho^2(e^2+e'^2)) \crm
&+& f_5\, \rho^2 e e' \cos(\omega+\omega') \crm
&+& f_6\, (e^4+8\rho^2 e'^2 \cos 2\omega') \crm
&+& f_7\, (e'^4+8\rho^2 e^2 \cos 2\omega) \crm
&+& f_8\, \rho^2 e e' \cos(\omega-\omega') \crm
&+& f_9\, e^3 e' \cos(\omega-\omega') \crm
&+& f_{10}\, e e'^3 \cos(\omega-\omega') \crm
&+& f_{11}\, e^2e'^2\cos 2(\omega-\omega') \crm
&+& f_{12}\, \rho^4\ ,
}
\label{eq.LeVerriera}
\ee
with $\rho = \sin (J/2)$. The secular Hamiltonian describing the evolution of a two planet system is then
\be
\bar{H}_{\rm close} = -\frac{{\cal G}mm'}{a'}
\moy{\lambda,\lambda'}{\frac{a'}{\Delta}}
\label{eq.LeVerrier}
\ee
In (\ref{eq.LeVerriera}), the orientation of the orbits are given relative to each other. More precisely, let G be the ascending node of the inner orbit relative to the outer one, and similarly, G' the ascending node of the outer orbit with respect to the inner one. G and G' are thus on the intersection between the two orbit planes, but in opposite directions (see Fig.~\ref{fig.LeVerrier}). The angles $\omega$ and $\omega'$ are the arguments of pericenter of the two planets relative to G and G', respectively.  With the notation of Fig.~\ref{fig.LeVerrier}, the longitudes of the ascending nodes of $m$ and $m'$ relative to the reference plane are $\Omega=\mathrm{ON}$ and $\Omega'=\mathrm{ON}'$, respectively.  Following \citet{LeVerrier_annales_1855}, we denote $\tau=\mathrm{ON}+\mathrm{N}\mathrm{G}$ and $\tau'=\mathrm{ON}'+\mathrm{N}'\mathrm{G}'$. The two arguments of pericenter $\omega$ and $\omega'$ are then given by
\figLeVerrier
\be
\EQM{
\omega  &= \varpi   - \tau  \cr
\omega' &= \varpi'  - \tau' \ .
}
\label{eq.peri}
\ee
Functions $(f_k)_{k=1,\ldots,12}$ contain the dependency in the semimajor axis ratio $\alpha=a/a'$. \citet{LeVerrier_annales_1855} computed them in terms of Laplace coefficients $\lap{s}{k}(\alpha)$, which are given by \citep[e.g.,][]{Laskar_Robutel_CeMDA_1995}
\be
\frac{1}{2}\lap{s}{k}(\alpha) = \frac{(s)_k}{k!} \alpha^k
F(s,s+k,k+1;\alpha^2)\ ,
\label{eq.Laplace}
\ee
and of their derivatives. In (\ref{eq.Laplace}), $(s)_0 = 1$ and $(s)_k=s(s+1)\cdots (s+k-1)$ if $k>0$. The function $F(a,b,c;x)$ is the Gauss hypergeometric function. Le Verrier's expressions of the functions $f_k$ are rather complicated. Here, we present instead those obtained using the algorithm described in \citet{Laskar_Robutel_CeMDA_1995} and implemented in the algebraic manipulator TRIP \citep{Gastineau_Laskar_trip_2012}:
\be
\EQM{
f_1 &=& \frac{1}{2}\lap{1/2}{0}\ , \crm
f_2 &=& \frac{1}{8}\alpha\lap{3/2}{1}\ , \crm
f_3 &=& -\frac{3}{4}\alpha\lap{3/2}{0}
       +\frac{1}{2}(1+\alpha^2)\lap{3/2}{1}\ , \crm
f_4 &=& \frac{9}{32}\alpha^2\lap{5/2}{0}\ , \crm
f_5 &=& \frac{9}{8}\alpha^2\lap{5/2}{1}\ , \crm
f_6 &=& -\frac{15}{128}\alpha^2\lap{5/2}{0}
       +\frac{3}{64}\alpha(1+3\alpha^2)\lap{5/2}{1}\ , \crm
f_7 &=& -\frac{15}{128}\alpha^2\lap{5/2}{0}
       +\frac{3}{64}\alpha(3+\alpha^2)\lap{5/2}{1}\ , \crm
f_8 &=& -\frac{15}{4}\alpha(1+\alpha^2)\lap{5/2}{0} \crm &&
       +\frac{3}{4}(2+3\alpha^2+2\alpha^4)\lap{5/2}{1}\ , \crm
f_9 &=& -\frac{15}{16}\alpha\lap{5/2}{0}
       +\frac{3}{32}(4+9\alpha^2)\lap{5/2}{1}\ , \crm
f_{10} &=& -\frac{15}{16}\alpha^3\lap{5/2}{0}
          +\frac{3}{32}\alpha^2(9+4\alpha^2)\lap{5/2}{1}\ , \crm
f_{11} &=& \frac{45}{64}\alpha^2\lap{5/2}{0}
       -\frac{9}{32}\alpha(1+\alpha^2)\lap{5/2}{1}\ , \crm
f_{12} &=& \frac{21}{8}\alpha^2\lap{5/2}{0}
       -\frac{3}{4}\alpha(1+\alpha^2)\lap{5/2}{1}\ .
}
\label{eq.funcLeVerrier}
\ee
In the averaged problem, the mean longitudes $(\lambda, \lambda')$ do not appear anymore, we thus discard their evolution. Furthermore, the semimajor axes are constant. The equations of motion derived by \citet{LeVerrier_annales_1855} of the other orbital elements of any two planets $m$ and $m'$ whatever are their position relative to each other, are
\be
\EQM{
\frac{de}{dt} &=& \frac{\sqrt{1-e^2}}{\Lambda e}\Dron{\bar H_{\rm close}}{\omega}\ , \crm
\frac{d\varpi}{dt} &=& -\frac{\sqrt{1-e^2}}{\Lambda e} \Dron{\bar H_{\rm close}}{e} 
                    + \tan\frac{I}{2}\sin I \frac{d\Omega}{dt}\ , \crm
\frac{dI}{dt} &=& \frac{\sin(\tau-\Omega)}{\Lambda \sqrt{1-e^2}}\Dron{\bar H_{\rm close}}{J}
\crm &&
                -\frac{\cos(\tau-\Omega)}{\Lambda\sqrt{1-e^2}}
\left(f_\tau-\frac{\rho}{\sqrt{1-\rho^2}} \Dron{\bar H_{\rm close}}{\omega}\right)\ , \crm
\frac{d\Omega}{dt} &=& -\frac{1}{\sin I}\Bigg(
                 \frac{\cos(\tau-\Omega)}{\Lambda\sqrt{1-e^2}} \Dron{\bar H_{\rm close}}{J}
\crm &&
                +\frac{\sin(\tau-\Omega)}{\Lambda\sqrt{1-e^2}}
\left(f_\tau-\frac{\rho}{\sqrt{1-\rho^2}} \Dron{\bar H_{\rm close}}{\omega}\right)\Bigg)
}
\label{eq.motionLeVerrier}
\ee
with
\be
f_\tau = \frac{1}{\sin
J}\left(\Dron{\bar H_{\rm close}}{\omega}+\Dron{\bar H_{\rm close}}{\omega'}\right)\ .
\ee
The relations between $(J,\tau)$ and the orbital elements relative to the reference plane are
\be
\EQM{
\cos J = \cos I \cos I' + \sin I \sin I' \cos(\Omega-\Omega'), \crm
\sin (\tau-\Omega) = \frac{\sin I'}{\sin J} \sin(\Omega-\Omega'), \crm
\cos (\tau-\Omega) = \frac{\cos I'-\cos I\cos J}{\sin I \sin J}\ .
}
\ee
The above equations are sufficient to compute numerically the secular evolution of a $p$-planet system with low mutual inclinations and low eccentricities. However, they have singularities when any inclination or eccentricity is nil. This issue can be circumvented by choosing non singular variables such as $k=e\cos\varpi$, $h=e\sin\varpi$, $q=\sin(I/2)\cos\Omega$, and $p=\sin(I/2)\sin\Omega$, and by imposing $\dot p = \dot q =0$ whenever $\sin J=0$. In any case, the equations of motion (\ref{eq.motionLeVerrier}) are complicated and do not facilitate the comprehension of the dynamical behavior of the system. For that reason, we shall consider a new set of variables.
\subsubsection{Milankovitch's variables}
\label{sec.Milankovitch}
The orientation in space and the shape of a Keplerian orbit (an ellipse) can be parametrized by two vectors: the dimensionless angular momentum $\vec j = \sqrt{1-e^2}\,\vec w$ and the Laplace-Runge-Lenz (or simply eccentricity) vector $\vec e = e\vec u$, where $\vec w$ is the unit vector normal to the orbit along the angular momentum, and $\vec u$ is the unit vector pointing toward the pericenter. These vectors are used in place of the usual elliptical elements $(e,\varpi,I,\Omega)$. Since each vector has three coordinates, the number of variables increases from four to six. This implies that the new variables are not independent, and indeed, they are related by two equations:
\be
\vec e \cdot \vec j = 0\ ,\quad
\norm{\vec e}^2 + \norm{\vec j}^2 = 1\ .
\ee
The equations of motion expressed in terms of these two vectors were derived by \citet{Milankovitch_Bull_1939}. In the secular problem, they read as \citep[e.g.,][]{Breiter_Ratajczak_MNRAS_2005, Tremaine_etal_AJ_2009, Rosengren_Scheeres_CeMDA_2014}
\be
\EQM{
\frac{d\vec j}{dt} &= -\frac{1}{\Lambda} 
( \vec j \times \grad{\vec j}\bar H + \vec e \times \grad{\vec e}\bar H)\ ,
\crm
\frac{d\vec e}{dt} &= -\frac{1}{\Lambda}
 (\vec e \times \grad{\vec j}\bar H + \vec j \times \grad{\vec e}\bar H)\ ,
}
\label{eq.Milankovitch}
\ee
where $\bar H$ (here $\bar H = \bar H_{\rm close}$) is the secular Hamiltonian of the system written in terms of the vectors ($\vec e, \vec j, \vec e', \vec j'$). The variables $(\vec e, \vec j)$ are not singular. Furthermore, they lead to more compact and symmetrical equations to describe the evolution of the system. The expression of the Hamiltonian as a function of these vectors is presented in the following section.
\subsubsection{Souriau's variables}
This section reproduces the computation of the perturbing function (\ref{eq.LeVerrier}) made by \citet{Abdullah_PhD_2001} in terms of the vectors $\vec e$ and $\vec j$\footnote{Our expressions differ slightly from those obtained by \citet{Abdullah_PhD_2001} because we use $\rho=\sin(J/2)$ like in the work of \citet{LeVerrier_annales_1855}, whereas in \citeauthor{Abdullah_PhD_2001}'s notation, $\rho=\sin J$.  Furthermore, the scaling factor in Laplace coefficients is arbitrary, and \citet{Abdullah_PhD_2001} do not include the factor $1/2$ that we have in (\ref{eq.Laplace}).}. This method involves new variables named after \citet{Souriau_book_1969}.  These variables noted $\vec \xi$ and $\vec \eta$, are defined by
\be
\EQM{
\vec \xi  = \vec j + \vec e\ , \crm
\vec \eta = \vec j - \vec e\ .
}
\label{eq.varSouriau}
\ee
It can easily be shown that
\be
\norm{\vec \xi} = 1\, 
\quad \norm{\vec \eta} = 1\ .
\label{eq.normSouriau}
\ee
Relations (\ref{eq.normSouriau}) let \citet{Souriau_book_1969} conclude that the set of ellipses with one fixed focus and semimajor axis is equivalent to the product of two spheres: $S^2\times S^2$. The dimension of the problem is thus clearly four. From (\ref{eq.Milankovitch}) and (\ref{eq.varSouriau}), the derivation of the equations of motion of $\vec \xi$ and $\vec \eta$ is straightforward. The result is
\be
\EQM{
\frac{d\vec \xi}{dt} = -
\frac{2}{\Lambda} \vec \xi \times \grad{\vec \xi}{H}\ , \crm
\frac{d\vec \eta}{dt} = -
\frac{2}{\Lambda} \vec \eta \times \grad{\vec \eta}{H}\ .
}
\label{eq.Souriau}
\ee
In these variables, the equations of motion (\ref{eq.Souriau}) are very simple and symmetric.
To get the expression of the secular Hamiltonian as a function of Souriau's variables, one needs to expand $\moy{}{a'/\Delta}$ in terms of these variables, but since their norms are equal to one, by construction, they are not small quantities. To solve that issue, \citet{Abdullah_PhD_2001} considered a new family of variables based on Souriau's one, and defined by
\be
\EQM{
T_1 &= \frac{1}{2}(1-\vec \xi \cdot \vec \eta)\ , \crm
T_2 &= \frac{1}{2}(1-\vec \xi' \cdot \vec \eta')\ ,\crm
V_1 &= \frac{1}{2}(1-\vec \xi \cdot \vec \xi')\ ,\crm
V_2 &= \frac{1}{2}(1-\vec \eta \cdot \vec \eta')\ , \crm
W_1 &= \frac{1}{2}(1-\vec \xi \cdot \vec \eta')\ ,\crm
W_2 &= \frac{1}{2}(1-\vec \xi' \cdot \vec \eta)\ .
}
\label{eq.Abdullah}
\ee
These variables can be interpreted as the square of the semi-distances between the points represented on the unit sphere by the vectors $\vec \xi$, $\vec \xi'$, $\vec \eta$, and $\vec \eta'$. Indeed, for instance,
\be
T_1 = \left(\frac{1}{2}\norm{\vec \xi - \vec \eta}\right)^2\ .
\ee
The variables (\ref{eq.Abdullah}) are quadratic in eccentricities and in mutual inclinations. The method is then to express all the quantities appearing in the expression (\ref{eq.LeVerrier}), i.e., $\rho^2$, $e^2$, $e'^2$, $p(k,\ell)$ and $q(k,\ell)$ as a function of the variables (\ref{eq.Abdullah}), where $p(k,\ell)$ and $q(k,\ell)$ are defined for all $(k,\ell)\in\mathbb{N}^2$ by
\be
\EQM{
p(k,\ell) = \rho^{k+\ell} e^k e'^\ell \cos(k\omega+\ell\omega')\ , \crm
q(k,\ell) = \rho^{\abs{k-\ell}} e^k e'^\ell \cos(k\omega-\ell\omega')\ ,
}
\label{eq.pqe}
\ee
if $k+\ell$ is even, and
\be
\EQM{
p(k,\ell) = \rho^{k+\ell} e^k e'^\ell \sin(k\omega+\ell\omega')\ , \crm
q(k,\ell) = \rho^{\abs{k-\ell}} e^k e'^\ell \sin(k\omega-\ell\omega')\ ,
}
\label{eq.pqo}
\ee
if $k+\ell$ is odd. To simplify the substitution, \citet{Abdullah_PhD_2001} provides recurrence relations to compute the $p(k,\ell)$ and $q(k,\ell)$ which are equivalent to those of Table~\ref{tab.recpq}.
\begin{table*}
\begin{center}
\caption{\label{tab.recpq} Recurrence relations for the computation of
$p(k,\ell)$ and $q(k,\ell)$, Eqs.~(\ref{eq.pqe})-(\ref{eq.pqo}).}
\renewcommand{\arraystretch}{1.4}
\begin{tabular}{ll} \hline
$p(k,\ell) = 2p(k-1,\ell-1)p(1,1) - \rho^4 e^2 e'^2 p(k-2,\ell-2)$ &
              $k \geq 2$ and $\ell \geq 2$ \\
$p(k,1) = 2p(k-1,0)p(1,1) - \rho^4 e^2 q(k-2,1)$ & $k>2$ \\
$p(1,\ell) = 2 p(0,\ell-1)p(1,1) + (-1)^\ell \rho^4 e'^2 q(1,\ell-2)$
             & $\ell>2$ \\
$p(k,0) = 2 (-1)^{k+1} p(k-1,0)p(1,0) + \rho^2 e^2 p(k-2,0)$ & $k \geq 2$ \\
$p(0,\ell) = 2 (-1)^{\ell+1} p(0,\ell-1)p(0,1) + \rho^2 e'^2 p(0,\ell-2)$
             & $\ell \geq 2$ \\
$p(2,1) = 2p(1,0)p(1,1) - \rho^2 e^2 q(0,1)$ & \\
$p(1,2) = 2p(0,1)p(1,1) + \rho^2 e'^2 q(1,0)$ & \\[0.5em]
$q{(k,\ell)} = 2q{(k-1,\ell-1)}q{(1,1)} - e^2 e'^2 q{(k-2,\ell-2)}$ & 
             $k \geq 2$ and $\ell \geq 2$ \\
$q{(k,1)} = 2q{(k-1,0)} q{(1,1)} - e^2 p{(k-2,1)}$ & $k\geq 2$ \\
$q(1,\ell) = 2q(0,\ell-1)q(1,1) + (-1)^\ell e'^2 p(1,\ell-2)$ &
             $\ell\geq2$ \\
$q(k,0) = p(k,0)\ ,\quad q(0,\ell) = (-1)^\ell p(0,\ell)$ & 
             $\forall k,l$ \\
\hline
\end{tabular}
\end{center}
\end{table*}
To initiate the recurrence, we have
\be
\EQM{
e^2  &= T_1\ , \crm
e'^2 &= T_2\ , \crm
\rho^2 &= \frac{1}{2}-\frac{2-V_1-V_2-W_1-W_2}{4\sqrt{(1-T_1)(1-T_2)}}\ , \crm
p(1,0) &= \frac{V_2-V_1+W_2-W_1}
                    {4\sqrt{(1-T_2)(1-\rho^2)}}\ , \crm
p(0,1) &= \frac{V_2-V_1+W_1-W_2}
                    {4\sqrt{(1-T_1)(1-\rho^2)}}\ , \crm
q(1,1) &= \frac{V_1+V_2-W_1-W_2}{2} \crm
& + 2 p(1,0) p(0,1)\ , \crm
p(1,1) &= \rho^2 \frac{V_1+V_2-W_1-W_2}{2}
\crm
& -2(1-\rho^2)p(1,0)p(0,1)\ ,
}
\label{eq.pqTVW}
\ee
and
\be
\EQM{
q(k,0) &= p(k,0)\ ,\crm
q(0,\ell) &= -p(0,\ell)
}
\ee
for all $(k,\ell)$ in $\mathbb{N}^2$.
After substituting the terms $p(k,\ell)$ and $q(k,\ell)$ in (\ref{eq.LeVerriera}) by $T_1$, $T_2$, $V_1$, $V_2$, $W_1$, and $W_2$, and truncating at the second order in these new variables, \citet{Abdullah_PhD_2001} obtained
\be
\EQM{
\moy{\lambda,\lambda'}{\frac{a'}{\Delta}} 
                   &= f_1 \crm
                   & +f_2 (2T_1+2T_2-V_1-V_2-W_1-W_2) \crm
                   & +\frac{1}{2}f_3 (V_1+V_2-W_1-W_2) \crm
                   & +g_4(\alpha^2 T_1^2 + T_2^2) \crm
                   & +g_5(V_1^2 + V_2^2) \crm
                   & +g_6(W_1^2 + W_2^2) \crm
                   & +g_7(T_1T_2+V_1V_2+W_1W_2) \crm
                   & +g_8(\alpha T_1-T_2)(V_1+V_2) \crm
                   & +g_9(\alpha T_1+T_2)(W_1+W_2) \crm
                   & +g_{10}(V_1W_1+V_2W_2) \crm
                   & +g_{11}(V_1W_2+V_2W_1)\ ,
}
\label{eq.HamAbdullah}
\ee
where $f_1$, $f_2$, and $f_3$ are given in (\ref{eq.funcLeVerrier}) and
\be
\EQM{
g_4 &= \frac{9}{32}\alpha \lap{5/2}{1}\ ,\crm
g_5 &= \frac{3}{16}\alpha(-5+4\alpha-5\alpha^2)\lap{5/2}{0}
      +\frac{3}{32}(2-\alpha+2\alpha^2)^2\lap{5/2}{1}\ , \crm
g_6 &= \frac{3}{16}\alpha(5+4\alpha+5\alpha^2)\lap{5/2}{0}
      -\frac{3}{32}(2+\alpha+2\alpha^2)^2\lap{5/2}{1}\ , \crm
g_7 &= \frac{9}{16}\alpha^2\lap{5/2}{0}\ ,\crm
g_8 &= -\frac{15}{32}\alpha(1-\alpha)\lap{5/2}{0}
       +\frac{3}{16}(1-\alpha^3)\lap{5/2}{1}\ ,\crm
g_9 &= -\frac{15}{32}\alpha(1+\alpha)\lap{5/2}{0}
       +\frac{3}{16}(1+\alpha^3)\lap{5/2}{1}\ ,\crm
g_{10} &= -\frac{3}{8}\alpha^2
        \left(\lap{5/2}{0}-\alpha\lap{5/2}{1}\right)\ , \crm
g_{11} &= -\frac{3}{8}\alpha
        \left(\alpha\lap{5/2}{0}-\lap{5/2}{1}\right)\ .
}
\label{eq.g_j}
\ee
The expansion (\ref{eq.HamAbdullah}) can be seen as a function of Souriau's variables $(\vec \xi, \vec \eta, \vec \xi', \vec \eta')$ using (\ref{eq.Abdullah}), or as a function of Milankovitch's variables $(\vec e, \vec j, \vec e', \vec j')$ using the definition of Souriau's variables (\ref{eq.varSouriau}). One gets
\be
\EQM{
\moy{\lambda,\lambda'}{\frac{a'}{\Delta}} 
&     =&c_1 \crm
&     +&c_2\,(e^2+e'^2+\vec j\cdot \vec j'-1) \crm
&     +&c_3 \,(\vec e \cdot \vec e') \crm
&     +&c_4 \,e^2e'^2 \crm
&     +&c_5 (\alpha^2 e^4 + e'^4) \crm
&     +&c_6 (\vec e \cdot \vec e')^2 \crm
&     +&c_7 (1-\vec j \cdot \vec j')^2 \crm
&     +&c_8 (\vec e \cdot \vec j')^2 \crm
&     +&c_9 (\vec j \cdot \vec e')^2 \crm
&     +&c_{10} \big(\alpha (1-\vec j\cdot \vec j') e^2 + (\vec e\cdot \vec e')e'^2\big) \crm
&     +&c_{11} \big((1-\vec j\cdot \vec j')e'^2 + \alpha (\vec e\cdot \vec e') e^2\big) \crm
&     +&c_{12} \big((1-\vec j\cdot \vec j')(\vec e\cdot \vec e') \crm
&      &   - (\vec e\cdot \vec j')(\vec j\cdot \vec e')\big)\ .
}
\label{eq.HamPP}
\ee
The first three lines with coefficients $c_1$, $c_2$, and $c_3$ correspond to the second order expansion in eccentricity and mutual inclination. All the other terms are associated to the order 4. Because the dot product of two vectors is invariant by any rotation applied on both vectors, the expression (\ref{eq.HamPP}) is clearly invariant by rotation. Thus, the reference frame does not need to be aligned with the mean plane of the planet system. The coefficients $(c_k)_{k=1,\cdots,12}$ are
\be
\EQM{
c_1 &= f_1\ , \crm
c_2 &= 2f_2\ , \crm
c_3 &= -f_3\ , \crm
c_4 &= g_7\ , \crm
c_5 &= g_4\ , \crm
c_6 &= \frac{1}{2}(g_5+g_6+g_7-g_{10}-g_{11})\ , \crm
c_7 &= \frac{1}{2}(g_5+g_6+g_7+g_{10}+g_{11})\ , \crm
c_8 &= \frac{1}{2}(g_5+g_6-g_7+g_{10}-g_{11})\ , \crm
c_9 &= \frac{1}{2}(g_5+g_6-g_7-g_{10}+g_{11})\ , \crm
c_{10} &= g_9+g_8\ , \crm
c_{11} &= g_9-g_8\ , \crm
c_{12} &= g_6-g_5\ .
}
\ee
Their explicit expressions in terms of Laplace coefficients are given in Tab.\ref{tab.ci}.
\Tabci
The equations of motion, derived from (\ref{eq.LeVerrier}), (\ref{eq.Milankovitch}), and (\ref{eq.HamPP}), are written in Appendix~\ref{app.compact}.
We stress here that the vectorial expansion (\ref{eq.HamPP}) is a generalization of the more standard development in eccentricity and absolute inclination \citep[e.g.,][]{Laskar_Robutel_CeMDA_1995, Ellis_Murray_Icarus_2000}. Indeed, at low inclination with respect to the reference frame, both formalisms are equivalent. If the planet system is tilted by a large angle, the linear equations of the Laplace-Lagrange approximation are still valid. But at higher orders, where inclinations get coupled with eccentricities, tilted systems can only be described with expansions in mutual inclination such as in the vectorial approach.  
\subsection{Hierarchical interaction}
Consider the case where the two planets $m$, $m'$ are very distant from each other ($\alpha \equiv a/a' \lesssim 0.1$).
The Hamiltonian $\bar H_{\rm hierar}$ governing the secular evolution is similar to $\bar H_{\rm close}$ (\ref{eq.LeVerrier}),
\be
\bar H_{\rm hierar} = -\frac{{\cal G}mm'}{a'} \moy{\lambda,\lambda'}{\frac{a'}{\Delta}}\ .
\label{eq.Hpcgen}
\ee
The only difference is that $\moy{}{a'/\Delta}$ is expanded in semimajor axis ratio $\alpha$ rather than in eccentricity and inclination. The most important contributions to this development have been made in the XIXth century \citep{Hansen_book_1853, Hill_analyst_1875, Tisserand_book_1889}. The explicit expression of the secular perturbing function expanded at the octupolar order is \citep[e.g.,][]{Laskar_Boue_AA_2010}\footnote{In this paper, $\omega'$ is defined with respect to the ascending node G' of the orbit $m'$ relative to the orbit $m$, while in \citet{Laskar_Boue_AA_2010}, $\omega'$ is defined with respect to the ascending node G of the orbit $m$ relative to the orbit $m'$. Thus, the two arguments of periastron differ by $\pi$.}
\be
\EQM{
\moy{\lambda,\lambda'}{\frac{a'}{\Delta}} &= 1
+ \Bigg( \left(\frac{1}{4}-\frac{3}{2}\rho^2+\frac{3}{2}\rho^4\right)
  \left(1+\frac{3}{2}e^2\right) 
\crm &+
  \frac{15}{4}\rho^2 \sigma^2 e^2\cos(2\omega)
\Bigg) \frac{\alpha^2}{(1-e'^2)^{3/2}} \crm
&+\frac{15}{16}\Bigg(\left(1+\frac{3}{4}e^2\right)
\crm & \times
\Big[ \sigma^2 \left(1-10\rho^2+15\rho^4\right)
 \cos(\omega-\omega') \crm & + 
\rho^2 \left(6-20\rho^2+15\rho^4\right) \cos(\omega+\omega')
\Big] \crm
&+ \frac{35}{4} e^2 \rho^2\sigma^2\Big[\rho^2\cos(3\omega+\omega')
\crm &
+\sigma^2  \cos(3\omega-\omega')\Big]
\Bigg) \frac{ee' \alpha^3}{(1-e'^2)^{5/2}}\ ,
}
\label{eq.Tisserand}
\ee
with $\sigma^2 = 1-\rho^2$.  As in section \ref{sec.planetplanet}, one can use the equations of motion (\ref{eq.motionLeVerrier}) derived by \citet{LeVerrier_annales_1855} to get the evolution of the system described by (\ref{eq.Tisserand}), but once again, the expressions are much simpler in a vectorial form. 
To get the vectorial expression of $\moy{}{a'/\Delta}$ expanded in semimajor axis ratio, we follow an algorithm very similar to the one of the section \ref{sec.planetplanet}. We define $P(k,\ell)$ and $Q(k,\ell)$, $(k,\ell)\in \mathbb{N}^2$, as
\be
\EQM{
P(k,\ell) &= \sigma^{\abs{k-\ell}} p(k,\ell)\ ,
\crm
Q(k,\ell) &= \sigma^{k+\ell}q(k,\ell)\ ,
}
\label{eq.PQ}
\ee
where $p(k,\ell)$ and $q(k,\ell)$ are given in Eqs~(\ref{eq.pqe})-(\ref{eq.pqo}). The secular part of the expansion in semimajor axis ratio of the perturbing function is a polynomial in $\alpha$, $\rho^2$, $e^2$, $e'^2$, $(1-e'^2)^{-1/2}$, $P(k,\ell)$, and $Q(k,\ell)$ \citep{Tisserand_book_1889}. The recurrence relations satisfied by $P(k,\ell)$ and $Q(k,\ell)$ are displayed in Table~\ref{tab.recPQ}.
\begin{table*}
\begin{center}
\caption{\label{tab.recPQ} Recurrence relations for the computation of
$P(k,\ell)$ and $Q(k,\ell)$, Eq.~(\ref{eq.PQ}).}
\renewcommand{\arraystretch}{1.4}
\begin{tabular}{ll} \hline
$P(k,\ell) = 2P(k-1,\ell-1)P(1,1) - \rho^4 e^2 e'^2 P(k-2,\ell-2)$ &
              $k \geq 2$ and $\ell \geq 2$ \\
$P(k,1) = 2P(k-1,0)P(1,1) - \rho^4 e^2 Q(k-2,1)$ & $k>2$ \\
$P(1,\ell) = 2 P(0,\ell-1)P(1,1) + (-1)^\ell \rho^4 e'^2 Q(1,\ell-2)$
             & $\ell>2$ \\
$P(k,0) = 2 (-1)^{k+1} P(k-1,0)P(1,0) + \sigma^2\rho^2 e^2 P(k-2,0)$ & $k \geq 2$ \\
$P(0,\ell) = 2 (-1)^{\ell+1} P(0,\ell-1)P(0,1) + \sigma^2\rho^2 e'^2 P(0,\ell-2)$
             & $\ell \geq 2$ \\
$P(2,1) = 2P(1,0)P(1,1) - \rho^2 e^2 Q(0,1)$ & \\
$P(1,2) = 2P(0,1)P(1,1) + \rho^2 e'^2 Q(1,0)$ & \\[0.5em]
$Q{(k,\ell)} = 2Q{(k-1,\ell-1)}Q{(1,1)} - \sigma^4 e^2 e'^2 Q{(k-2,\ell-2)}$ & 
             $k > 2$ and $\ell \geq 2$ \\
$Q{(k,1)} = 2Q{(k-1,0)} Q{(1,1)} - \sigma^4 e^2 P{(k-2,1)}$ & $k> 2$ \\
$Q(1,\ell) = 2Q(0,\ell-1)Q(1,1) + (-1)^\ell \sigma^4 e'^2 P(1,\ell-2)$ &
             $\ell>2$ \\
$Q(k,0) = P(k,0)\ ,\quad Q(0,\ell) = (-1)^\ell P(0,\ell)$ & 
             $\forall k,l$ \\
$Q(2,1) = 2Q(1,0)Q(1,1) - \sigma^2 e^2 P(0,1)$ & \\
$Q(1,2) = 2Q(0,1)Q(1,1) + \sigma^2 e'^2 P(1,0)$ & \\
\hline
\end{tabular}
\end{center}
\end{table*}
The initialization of the recurrence can be done either in terms of Souriau's variables
\be
\EQM{
e^2 &=& T_1\ , \crm
e'^2 &=& T_2\ ,\crm
\rho^2 &=& \frac{1}{2} -
      \frac{2-V_1-V_2-W_1-W_2}{4\sqrt{(1-T_1)(1-T_2)}}\ , \crm
P(1,0) &=& \frac{V_2-V_1+W_2-W_1}{4\sqrt{1-T_2}}\ , \crm
P(0,1) &=& \frac{V_2-V_1+W_1-W_2}{4\sqrt{1-T_1}}\ , \crm
Q(1,1) &=& \sigma^2\frac{V_1+V_2-W_1-W_2}{2} \crm
&&+2P(1,0)P(0,1) \ ,\crm
P(1,1) &=& \rho^2\frac{V_1+V_2-W_1-W_2}{2} \crm
&&- 2 P(1,0)P(0,1)\ ,
}
\label{eq.SouriauPQ}
\ee
or directly in terms of Milankovitch's vectors
\be
\EQM{
2 \rho^2   &=& 1-\vec w \cdot \vec w'\ , \crm
2 \sigma^2 &=& 1+\vec w \cdot \vec w'\ , \crm
2 P(1,0) &=& (\vec w' \cdot \vec e)\ , \crm
2 P(0,1) &=& (\vec w \cdot \vec e')\ , \crm
2 Q(1,1) &=& -2\sigma^2(\vec e\cdot \vec e') 
           +(\vec e\cdot\vec w')(\vec w\cdot\vec e')\ ,\crm
2 P(1,1) &=& -2\rho^2(\vec e\cdot\vec e')
           -(\vec e\cdot\vec w')(\vec w\cdot\vec e')\ ,\crm
}
\label{eq.MilankovitchPQ}
\ee
where $\vec w=\vec j/\sqrt{1-e^2}$, and $\vec w'=\vec j'/\sqrt{1-e'^2}$.  A quick comparison of (\ref{eq.SouriauPQ}) and (\ref{eq.MilankovitchPQ}) suggests that Milankovitch's formalism is more adapted for this problem.  In these variables $\moy{}{a'/\Delta}$, expanded in semimajor ratio, reads
\be
\EQM{
\moy{\lambda,\lambda'}{\frac{a'}{\Delta}} &= 1 
+\frac{\alpha^2}{8j'^5} \Big(3(\vec j\cdot\vec j')^2
-(1-6e^2)j'^2
\crm &
-15(\vec e\cdot\vec j')^2\Big) 
+\frac{15\alpha^3}{64j'^7} \Big(
(\vec e\cdot \vec e')
\crm & \times
\left[(1-8e^2)j'^2+35(\vec e\cdot\vec j')^2 -5(\vec j\cdot\vec j')^2\right]
\crm &
-10(\vec e\cdot\vec j')(\vec j\cdot\vec e')(\vec j\cdot\vec j')
\Big) \ .
}
\label{eq.HamPC}
\ee
One can easily check that (\ref{eq.HamPC}) is invariant by rotation.  The associated equations of motion of the planet and the companion are deduced from (\ref{eq.Milankovitch}), (\ref{eq.Hpcgen}), and (\ref{eq.HamPC}) (see Appendix~\ref{app.hierarchical}).
\subsection{Spin-orbit interaction}
Because of their proper rotations, stars are not spherical and exert a torque on the orbital motion of their planets. Let a system composed of a star $m_0$ with one planet $m$. We note $\vec s$ the unit vector along the spin axis of the star, $J_2$ its quadrupole gravitational harmonic, and $R_0$ its equatorial radius. We make the assumption that $\vec s$ also corresponds to the stellar axis of maximal inertia (gyroscopic approximation). The secular quadrupole potential energy due to the oblateness of the star acting on the planet is \citepalias[e.g.,][]{Boue_Laskar_Icarus_2006}
\be
\bar H_{\rm spin} = \frac{{\cal G} m_0 m J_2 R_0^2}{4 a^3 (1-e^2)^{3/2}}
\left(1-3(\vec s\cdot \vec w)^2\right)\ .
\label{eq.HamSP}
\ee
This expression is valid as long as the distance of the planet to the star is much larger than the stellar radius ($r\gg R_0$).  Nevertheless, since stellar deformations are usually small, we assume that (\ref{eq.HamSP}) is valid even for close-in planets with very small semimajor axis. There is no assumption regarding the obliquity of the star relative to the orbital plane of the planet.
In a generic problem where $\bar H$ represents the Hamiltonian of the system, the equation of motion satisfied by the spin axis is \citepalias{Boue_Laskar_Icarus_2006}
\be
\frac{d\vec s}{dt} = -\frac{1}{L} \vec s \times \grad{\vec s}{\bar H}\ ,
\label{eq.dsdt}
\ee
where $\vec L=C \omega_0\vec s$ is the angular momentum of the star, $C$ its moment of inertia along the $\vec s$ axis, and $\omega_0$ its rotation rate. The explicit equations of motion of $\vec s$, $\vec j$ and $\vec e$ are given in Appendix~\ref{app.spinorbit}.  As shown in \citetalias{Boue_Laskar_Icarus_2006}, the kinetic energy associated to the rotation of a rigid body around its spin axis, which should be added in the Hamiltonian, does not contribute to the equations of motion once the Hamiltonian is averaged over the proper rotation of the rigid body, or when the body has an axial symmetry. As a consequence, we drop this kinetic energy from our equations.
The gravity field coefficient $J_2$ is deduced from the rotation speed of the star according to \citep[e.g.,][]{Lambeck_book_1988}
\be
J_2 = k_2 \frac{\omega_0^2R_0^3}{3 {\cal G} m_0}\ ,
\label{eq.Jom}
\ee
where $k_2$ is the second Love number.
\subsection{Relativistic precession}
The last effect taken into account is the secular contribution of general relativity on the precession motion of the planets pericenter.  For a planet $m$ orbiting a star $m_0$, the associated Hamiltonian reads \citep[e.g.,][]{Touma_etal_MNRAS_2009}
\be
\bar H_{\rm relat} = -\frac{3\mu^2\beta}{a^2c^2\sqrt{1-e^2}}\ ,
\label{eq.HamRG}
\ee
where $c$ is the speed of light, $\beta = m_0m/(m_0+m)$, and $\mu = {\cal G}(m_0+m)$. The associated equations of motion are given in Appendix~\ref{app.genrel}.
\subsection{Example}
The formalism described above can be used to model many different types of systems. For example, consider a
compact planet system perturbed by an outer stellar companion such as the 55~Cancri system \citep{Kaib_etal_ApJ_2011}. If the planet system is similar to those detected by the {\em Kepler} spacecraft, it should be dynamically cold with low eccentricities and mutual inclinations. We further assume that the dynamics of the system is not dominated by mean-motion resonances. Conversely, the binary component can be highly eccentric with large inclination with respect to the planet plane. In this case, the Hamiltonian of the system can be approximated by
\be
\EQM{
H_{\rm tot} =& \sum_{1\leq j<k\leq p}\bar{H}_{\rm close}(j,k) + 
              \sum_{j=1}^{p} \Big(\bar{H}_{\rm hierar}(j,p+1)
             \crm &
             + \bar{H}_{\rm spin}(0,j) + \bar{H}_{\rm relat}(j)\Big)\ ,
}
\label{eq.Hamtot}
\ee
where $p$ is the number of planets, and $\bar H_{\rm close}(j,k)$ (\ref{eq.HamPP}), $\bar H_{\rm hierar}(j,p+1)$ (\ref{eq.HamPC}), $\bar H_{\rm spin}(0,j)$ (\ref{eq.HamSP}), and $\bar H_{\rm relat}(j)$ (\ref{eq.HamRG}) represent the interactions between planets $j$ and $k$, the interaction between the planet $j$ and the companion, the spin-orbit interaction between the star and the planet $j$, and the relativistic precession of planet $j$, respectively. Naturally, this model is only valid as long as the eccentricity of each planet and the mutual inclination of any pairs of planets remain low. But an increase of any of these two quantities would be the signature of instability and would thus be informative as well.
\section{Numerical tests and applications}
\label{sec.numerical}
To check our secular models, we performed several tests. First, we considered a system composed of the four inner planets of our solar system. We made this choice because the eccentricities and the mutual inclinations are well known, and relatively low. We used the initial conditions provided in \citet{Yoder_geph.conf_1995}. This system is used to test the expansions (\ref{eq.LeVerrier}) and (\ref{eq.HamPP}) of the perturbing function in small eccentricity and mutual inclination. In a second step, we chose a hierarchical system undergoing Lidov-Kozai oscillations \citep{Lidov_PSS_1962, Kozai_AJ_1962} to test our expansions in semimajor axis (\ref{eq.Tisserand}) and (\ref{eq.HamPC}). In both cases, the secular evolutions are compared to numerical simulations done with a full $n$-body symplectic integrator. 
\subsection{A compact system}
Figure~\ref{fig.comparison1} displays the evolution over one million years of the eccentricities and inclinations of the four planets of our first system with respect to the initial ecliptic plane ($I_{\rm Earth}=0$ at $t=0$) obtained with an $n$-body integrator and by solving the Hamiltonian (\ref{eq.HamPP}). General relativity is included in both simulations as is the effect of the solar oblateness. The two integrations give very similar results and their distinction is hardly perceptible in most subfigures.  It is best seen in the evolution of the eccentricity of Venus and of the Earth, where it consists mostly in a slight shift in the precession frequencies. Thus, the secular formulation preserves the main dynamical features of this system, as already observed by, e.g., \citet{Laskar_Icarus_2008}. The integration of the Hamiltonian (\ref{eq.LeVerrier}) using Le~Verrier's equations of motion is not shown here, because it cannot be distinguished from the solution obtained with the vectorial approach.
\figCompa
\figCompb
\figCompbb
In Fig.~\ref{fig.comparison2} and \ref{fig.comparison2b}, we test the effect of a rotation of the whole system by an angle $\Delta I\in \{$0, 10, 60, 90, 135, 180$\}$ degrees.  More precisely, we apply each of these rotations on the secular integrations only, and we keep the $n$-body integration of the previous figure unchanged.  Since the choice of the reference frame is arbitrary, the eccentricities should not be affected by these rotations. Figure~\ref{fig.comparison2} shows that this is indeed the case whether the evolution is obtained using Le~Verrier's equations (\ref{eq.motionLeVerrier}) or with the vectorial approach (\ref{eq.Milankovitch}). 
On the other hand, Hamiltonians expanded in {\em absolute} inclinations are not designed to describe the evolution of such highly inclined systems. It is thus natural to observe discrepancies above $\Delta I=90^\circ$ between the secular model taken from, e.g., \citet{Laskar_Robutel_CeMDA_1995} (dotted curve in Fig.~\ref{fig.comparison2}) and the $n$-body integration (solid curve).
A similar result is observed on Mercury's inclination in Fig.~\ref{fig.comparison2b}.  To make this figure, we integrated the system in a tilted frame, and then we applied a rotation of $-\Delta I$ on the output to place the system back in the reference frame of the $n$-body integration. The lack of precision with the expansion in absolute inclination was expected since absolute inclinations are not small as $\Delta I$ increases.  However, it is interesting to see that even at $\Delta I=60^\circ$, the Hamiltonian expanded in absolute inclination provides reasonable evolutions. This is due to the fact that the system is well described by the Laplace-Lagrange approximation (second order in inclination and eccentricity), and that this approximation is invariant by rotation.  Once again, Le~Verrier's formalism and the vectorial approach match perfectly, and remain in very good agreement with the full $n$-body integration.
\subsection{Lidov-Kozai oscillations}
\figCompc
To test our expansions in semimajor axis (\ref{eq.Tisserand}) and (\ref{eq.HamPC}), we integrate a system composed of a planet with mass $m_1=1 M_J$, semimajor axis $a_1 = 6$ au, and initial eccentricity $e_1=0.001$ and a brown dwarf with mass $m_2=40 M_J$, semimajor axis $a_2=100$ au, and initial eccentricity $e_2 = 0.6$. The mass of the central star is $m_0=1M_\odot$, and the mutual inclination is initially set to $J=65^\circ$. With these values, the system undergoes Lidov-Kozai oscillations whose modeling requires the octupole order \citep{Naoz_etal_nature_2011}. The integrations are performed with an $n$-body code, and with the secular approximations (\ref{eq.Tisserand}) and (\ref{eq.HamPC}). In all simulations general relavity (\ref{eq.HamRG}) is included. The results are displayed Fig.~\ref{fig.comparison3}. 
The two secular models are strictly equivalent, the solutions are thus indistinguishable. This was not the case for the Hamiltonian expanded in mutual inclination and eccentricities since the quantities $p(k,\ell)$ and $q(k,\ell)$ Eq.~(\ref{eq.pqTVW}), present in the Hamiltonian (\ref{eq.LeVerrier}), were themselves expanded in $T_1$, $T_2$, $V_1$, $V_2$, $W_1$, and $W_2$ to get the expression (\ref{eq.HamPP}). The comparison with the $n$-body integration is also very good. The integrations differ slightly once the planet flips into a retrograde orbit. This transition does not occur at the exact same time in the $n$-body and the secular simulations, as a result, the direction of the ninth kick in the evolution of $I_1$ is not the same in the two integrations. Nevertheless, the transition is preceded by a passage through an extreme and unrealistic eccentricity where the periastron distance, 0.06$R_\odot$, is much shorter than the size of the central star.
\section{Conclusion}
We have provided a vectorial formalism for studying the secular motion of non-resonant conservative gravitational systems with concentric orbits such as planetary systems. All expressions are analytical and expressed in terms of the vectors $\vec e$ and $\vec j$, namely the Laplace-Runge-Lenz vector and the dimensionless orbital angular momentum. Planet-planet interactions have been developed either in eccentricity and mutual inclination or in semimajor axis ratio 
using an algorithm adapted from \citet{Abdullah_PhD_2001}.
For completeness, the vectorial expression of the spin-orbit interaction and general relativity have also been recalled.
With numerical tests, we have shown that the integrations of the vectorial equations are in perfect agreement with the more standard approach relying on classical elliptic elements $(a,e,I,\lambda,\varpi,\Omega)$.
The \emph{Kepler} spacecraft has revealed a population of low-inclination, low-mass, and low-period planets which are not readily studied with N-body techniques.  The present formalism can be used to study their dynamics, including the onset of secular chaos.  
Moreover, the vectorial approach naturally is expanded to include a companion on a hierarchical orbit.  That companion could be a binary star companion, or a giant planet at a considerable distance.  As long as the inner planet system remains with coplanar orbits and low eccentricities, our formalism remains valid.  It can, in fact, be used to see whether the system will remain stable, or whether secular chaos will lead to orbit crossings, which then must be followed with an N-body approach.  
Inclusion of the precession dynamics of central star also allows us to address the problem of spin-orbit alignment in multiplanet systems, which is newly observationally relevant, and which we will pursue in a follow-up paper (Bou\'e and Fabrycky, 2014). 
\acknowledgments
GB thanks Philippe Robutel, Jacques Laskar and Alain Albouy for the many discussions which have been helpful for this study, and in particular those about Khaled Adbulah's PhD thesis.
\appendix
\section{Explicit vectorial equations of motion}
In this appendix, we provide the explicit expressions of the secular equations of motion written in a vectorial form. Notations are the same as in Section~\ref{sec.Milankovitch}. We consider both compact and hierarchical systems. In each case, the parameters of the outer body are noted with a prime, while those of the inner body are unprimed.
\subsection{Compact system}
\label{app.compact}
The Hamiltonian (\ref{eq.HamPP}) describing the secular evolution of a compact two planet system reads
\be
\EQM{
H_1 =& -\frac{{\cal G}mm'}{a'} \Big[
  c_1 
+ c_2 (e^2+e'^2+\vec j\cdot \vec j' -1)
+ c_3(\vec e\cdot \vec e')
+ c_4 e^2e'^2
+ c_5(\alpha^2 e^4+e'^4) 
+ c_6(\vec e\cdot\vec e')^2 
\crm &
+ c_7(1-\vec j\cdot\vec j')^2
+ c_8(\vec e\cdot\vec j')^2 
+ c_9(\vec j\cdot\vec e')^2
+ c_{10}\left(\alpha(1-\vec j\cdot\vec j')e^2+(\vec e\cdot\vec e')e'^2\right)
\crm &
+c_{11}\left((1-\vec j\cdot\vec j')e'^2+\alpha(\vec e\cdot\vec e')e^2\right)
+c_{12}\left((1-\vec j\cdot\vec j')(\vec e\cdot\vec e')-(\vec e\cdot\vec
j')(\vec j\cdot\vec e')\right)
\Big]\ ,
}
\ee
where the $c_i$'s are parameters depending on Laplace coefficients (see Tab.~\ref{tab.ci}). The conservation of the orbital angular momentum implies that
\be
\EQM{
\Lambda \frac{d\vec j}{dt} = -\Lambda' \frac{d\vec j'}{dt'} = \vec T_1\ , 
}
\ee
where $\vec T_1 = -\vec j\times \grad{\vec j}{H_1} - \vec e\times
\grad{\vec e}{H_1}$ is a torque whose expression is
\be
\vec T_1 = \frac{{\cal G} m m'}{a'} \Big(
  A_1\,\vec j\times\vec j' 
+ B_1\,\vec e\times\vec e'
+ C_1\,\vec e\times\vec j'
+ D_1\,\vec j\times\vec e'
\Big)\ .
\ee
The other equations of motion, deduced from (\ref{eq.Milankovitch}), are
\be
\frac{d\vec e}{dt} = \frac{{\cal G}mm'}{a'\Lambda} \Big(
  A_1\,\vec e\times\vec j'
+ B_1\,\vec j\times\vec e'
+ C_1\,\vec j\times\vec j'
+ D_1\,\vec e\times\vec e'
+ E_1\,\vec j\times\vec e
\Big)\ ,
\ee
and
\be
\frac{d\vec e'}{dt} = \frac{{\cal G}mm'}{a'\Lambda'} \Big(
  A_1\,\vec e'\times\vec j
+ B_1\,\vec j'\times\vec e
+ C_1\,\vec e'\times\vec e
+ D_1\,\vec j'\times\vec j
+ F_1\,\vec j'\times\vec e'
\Big)\ ,
\ee
with
\be
\EQM{
A_1 &= c_2 - 2c_7(1-\vec j\cdot \vec j') - \alpha c_{10} e^2 -c_{11}e'^2 - c_{12}(\vec e\cdot\vec e')\ ,\crm
B_1 &= c_3 + 2c_6(\vec e\cdot\vec e') + c_{10} e'^2 + \alpha c_{11}e^2 + c_{12}(1-\vec j\cdot\vec j')\ ,\crm
C_1 &= 2 c_8 (\vec e\cdot \vec j') - c_{12}(\vec j\cdot\vec e')\ ,\crm
D_1 &= 2 c_9 (\vec j\cdot \vec e') - c_{12}(\vec e\cdot\vec j')\ ,\crm
E_1 &= 2c_2 + 2c_4 e'^2 + 4\alpha^2 c_5 e^2 + 2\alpha c_{10}(1-\vec j\cdot\vec j') + 2\alpha c_{11}(\vec e\cdot \vec e')\ ,\crm
F_1 &= 2c_2 + 2c_4 e^2 + 4c_5 e'^2 + 2c_{10}(\vec e\cdot\vec e') + 2c_{11}(1-\vec j\cdot\vec j')\ .
}
\ee
\subsection{Hierarchical system}
\label{app.hierarchical}
The Hamiltonian (\ref{eq.HamPC}) describing the secular evolution of a hierarchical two planet system reads
\be
\EQM{
H_2 =& -\frac{{\cal G}mm'}{a'} \Big[
1 + \frac{\alpha^2}{8j'^5}\left(3(\vec j\cdot\vec j')^2 - (1-6e^2)j'^2 - 15 (\vec e\cdot\vec j')^2\right)
\crm &
+ \frac{15 \alpha^3}{64 j'^7} \Big(
\big[(1-8e^2)j'^2 + 35(\vec e\cdot\vec j')^2 - 5(\vec j\cdot\vec j')^2\big](\vec e\cdot\vec e')
- 10(\vec e\cdot\vec j')(\vec j\cdot\vec e')(\vec j\cdot\vec j')
\Big)
\Big]\ , 
}
\ee
where $j=\norm{\vec j} = \sqrt{1-e^2}$.  The conservation of the orbital angular momentum implies that
\be
\Lambda \frac{d\vec j}{dt} = -\Lambda' \frac{d\vec j'}{dt} = \vec T_2\ ,
\ee
where $\vec T_2 = -\vec j\times \grad{\vec j}{H_2} - \vec e\times \grad{\vec e}{H_2}$ is a torque whose expression is
\be
\vec T_2 = \frac{{\cal G} m m'}{a'} \Big(
  A_2\,\vec j\times\vec j' 
+ B_2\,\vec e\times\vec e'
+ C_2\,\vec e\times\vec j'
+ D_2\,\vec j\times\vec e'
\Big)\ .
\ee
The other equations of motion, deduced from (\ref{eq.Milankovitch}), are
\be
\frac{d\vec e}{dt} = \frac{{\cal G}mm'}{a'\Lambda} \Big(
  A_2\,\vec e\times\vec j'
+ B_2\,\vec j\times\vec e'
+ C_2\,\vec j\times\vec j'
+ D_2\,\vec e\times\vec e'
+ E_2\,\vec j\times\vec e
\Big)\ ,
\ee
and
\be
\frac{d\vec e'}{dt} = \frac{{\cal G}mm'}{a'\Lambda'} \Big(
  A_2\,\vec e'\times\vec j
+ B_2\,\vec j'\times\vec e
+ C_2\,\vec e'\times\vec e
+ D_2\,\vec j'\times\vec j
+ F_2\,\vec j'\times\vec e'
\Big)\ ,
\ee
with
\be
\EQM{
A_2 &=& \frac{3\alpha^2}{4j'^5}(\vec j\cdot\vec j') 
     - \frac{75\alpha^3}{32j'^7}\left[(\vec e\cdot\vec j')(\vec
j\cdot\vec e') + (\vec e\cdot\vec e')(\vec j\cdot \vec j')\right]\ , \crm
B_2 &=& \frac{15\alpha^3}{64j'^7}
\left[(1-8e^2)j'^2+35(\vec e\cdot\vec j')^2-5(\vec j\cdot\vec j')^2\right]\ ,\crm
C_2 &=& -\frac{15\alpha^2}{4j'^5} (\vec e\cdot\vec j')
      + \frac{75\alpha^3}{32j'^7} \left[7(\vec e\cdot\vec j')(\vec
e\cdot \vec e') - (\vec j\cdot\vec e')(\vec j\cdot\vec j')\right]\ ,\crm
D_2 &=&-\frac{75\alpha^3}{32j'^7}(\vec e\cdot\vec j')(\vec j\cdot\vec j')\ ,\crm
E_2 &=& \frac{3\alpha^2}{2j'^3} - \frac{15\alpha^3}{4j'^5}(\vec e\cdot\vec e')\ ,\crm
F_2 &=& \frac{3\alpha^2}{8j'^7}\left[5(\vec j\cdot\vec j')^2 - (1-6e^2)j'^2-25(\vec e\cdot\vec j')^2\right]
\crm &&
+ \frac{75\alpha^3}{64j'^9}\Big(\left[(1-8e^2)j'^2+49(\vec e\cdot\vec j')^2
-7(\vec j\cdot\vec j')^2\right](\vec e\cdot \vec e')
-14(\vec e\cdot\vec j')(\vec j\cdot\vec e')(\vec j\cdot\vec j')
\Big)\ .
}
\ee
\subsection{Spin-orbit interaction}
\label{app.spinorbit}
The Hamiltonian (\ref{eq.HamSP}) governing the spin-orbit evolution of a planet orbiting an oblate star is
\be
H_3 = \frac{Gm_0mJ_2R_0^2}{4a^3(1-e^2)^{3/2}}\left(1-3(\vec s\cdot\vec w)^2\right)\ .
\ee
Once again, the conservation of the angular momentum implies that
\be
L \frac{d\vec s}{dt} = - \Lambda\frac{d\vec j}{dt} = \vec T_3\ ,
\ee
with $\vec T_3 = -\vec s \times \grad{\vec s} H_3$ is the torque acting on the stellar rotation. Its expression is
\be
\vec T_3 = \frac{3}{2}\frac{Gm_0mJ_2R_0^2}{a^3(1-e^2)^{3/2}}
\left(\vec s\cdot\vec w\right)\vec s\times \vec w\ .
\ee
The evolution of the eccentricity vector $\vec e$, deduced from (\ref{eq.Milankovitch}), is given by
\be
\Lambda\sqrt{1-e^2} \frac{d\vec e}{dt} = -\frac{3}{2}\frac{Gm_0mJ_2R_0^2}{a^3(1-e^2)^{3/2}}
\left[\left(\vec s\cdot\vec w\right)\vec s\times \vec e
+\frac{1}{2}\left(1-5(\vec s\cdot\vec w)^2\right)\vec w\times \vec e
\right]\ .
\ee
\subsection{General relativity}
\label{app.genrel}
The effect of general relativity induced by the massive central star is modeled by the Hamiltonian (\ref{eq.HamRG})
\be
H_4 = -\frac{3\mu^2\beta}{a^2c^2\sqrt{1-e^2}}\ .
\ee
The orbital angular momentum of the planet is conserved, and its eccentricity vector evolves according to
\be
\frac{d\vec e}{dt} = \frac{3\mu^2\beta}{a^2c^2\Lambda(1-e^2)}\vec
w\times\vec e\ .
\ee
\bibliographystyle{apj}
\bibliography{bf}
\end{document}